# BRG.LifeMOD™ modeling and simulation of swimmers impulse during a grab start

Guillaume Agnesina[1], Redha Taiar[1], Nicolas Houel[2], Kevin Guelton[3], Philippe Hellard[2], Yuli Toshev[1, 4]

[1] Laboratory for Analysis of Mechanical Constraints LRC/CEA EA 3304, University of Reims CA, France
[2] FFN, INSEP, Paris, France
[3] CRESTIC, University of Reims CA, France
[4] Institute of Mechanics and Biomechanics, Bulgarian Academy of Sciences, Sofia, Bulgaria

*Abstract*—-The main aim of the study is to propose an approach for 3D modeling and simulation of swimmers impulse during a grab start. Four national level swimmers were investigated. For every swimmer, 3D model was generated in BRG.LifeMOD™ on the base of individual morphologic parameters and kinematic data obtained by high speed video camera and passive markers attached at the level of each important articulation. The proposed approach allows predicting swimmer's joint moments for each important articulation during the impulse phase of the grab start and to analyze the segmental coordination for each studied swimmer in order to optimize the performance. The model was successfully validated by comparing the predicted speed and power values with experimental ground reactions data collected in situ.

## I. INTRODUCTION

Whatever the swimming disciplines the study of swimmers' performance involves the identification of three technical phases: start, turn and strokes phases. Analysis of the temporal distribution showed that the start phase accounts for 15 % and 7.7 % of total time, respectively for 50 m and 100 m freestyle events. In short distance races (50m and 100m) the start represents a particularly important factor. For example, at the Athens Olympic Games (2004), the time separating the eight finalists in the men's 50m freestyle final was 0.44 s, which represents 2% of the winner's total race time (21.93 s). The difference in performance among the finalists' may find an explanation in the time lost during the start phase.

Regardless of underwater factors, the start phase depends primarily on the quality of the swimmer's impulse on the starting platform (Vilas-Boas et al., 2003). Although recent studies have been undertaken, using both dynamics and kinematics approaches, they do not yield additional information concerning the relationship between the swimmers' movements and their actual performance. Few studies have addressed the modeling (dynamic and/or kinematics) of the parameters that determine the performance according to swimmers' movement during start phases (Holtes and McLean, 2001). It is to note that modeling methods used for movements study in others sports, for example skiing (Houel, 2004), seem to be the most effective for detailed understanding of movements and for performance prediction.

The main aim of the study is to develop an approach for 3D modeling and simulation of swimmers impulse during a grab start in order to optimize the performance.

## II. METHODS

Four national level swimmers were instructed to perform a grab start. Subjects' average height and mass were respectively 183.5 cm (± 3.41) and 75.77 kg (± 3.89). Swimmers were equipped with 16 passive markers (8 on the right and 8 on the left side) attached at the level of each important articulation: foot, ankle, knee, hip, shoulder, elbow, wrist, finger (Fig.1).

For each start, 3 high speed video cameras (125 Hz) were used. The most important data were obtained by the two cameras in the two sagittal (lateral) planes (right and left). Simultaneously, the ground reaction forces were measured using a 3D force-plate AMTI OR6-7-2000 mounted on the start block (Fig. 1) with sampling frequency of 1000 Hz.

Swimmer's segments centres of mass velocity were calculated by integration of acceleration values. Before each start, the video cameras and the force-plate were calibrated and clock synchronized (0.008 s accuracy).

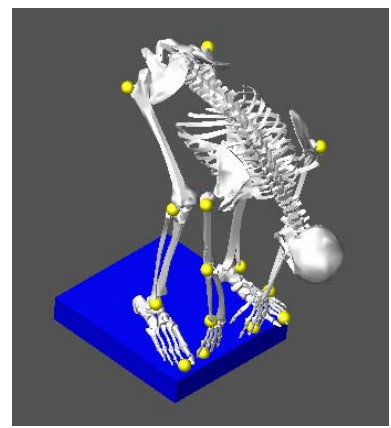

Figure 1. Passive markers (8 on the right and 8 on the left side) were attached at the level of each important articulation. AMTI force-plate was mounted on the block.

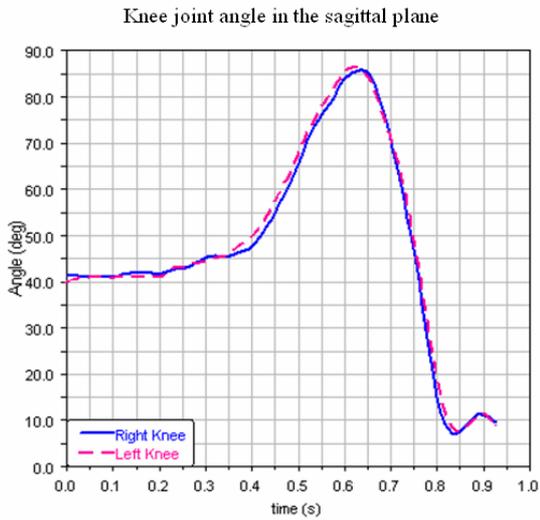

Figure 2. The swimmer's knee joint angles (right and left knee joint) obtained using high sped video cameras and passive maerkers and after that used as input data for the 3D swimmer body model.
Time period: 0.00s ÷ 0.92s

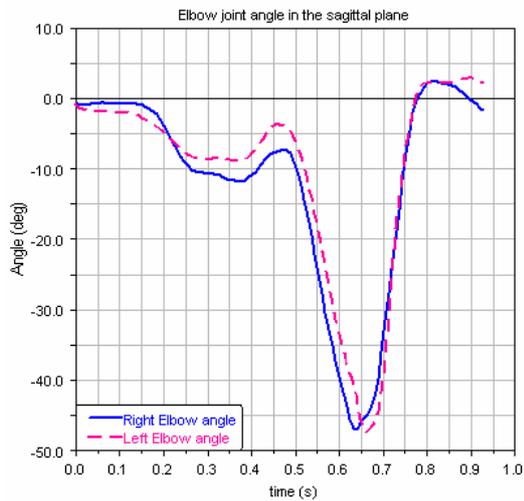

Figure 3. The swimmer's elbow joint angles (right and left knee elbow) obtained using high sped video cameras and passive markers and after that used as input data for the 3D swimmer body model.
Time period = 0.00s ÷ 0.92s

Two-dimensional video analysis was carried out for the impulse phase, i.e. during the period when the athletes were in contact with the force-plate mounted on the start block. The video analysis allowed determining the angles between the subjects' segments axes and the horizontal plane. The data were fitted using polynomials' method (Tavernier et al, 1996; Winter et al, 1990).

The swimmer's knee and elbow joint angles (right and left knee elbow) were obtained using high sped video cameras and passive markers and are shown on Fig. 2 and Fig. 3. After that they were used as input data for the 3D modeling.

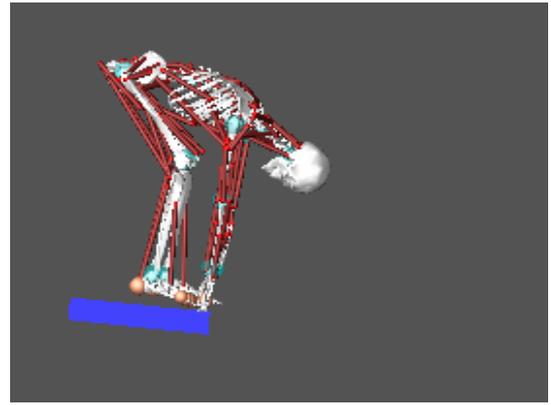

Figure 4. 3D suimmer's body model in the initial moment of the impulse phase. Time = 0 s;

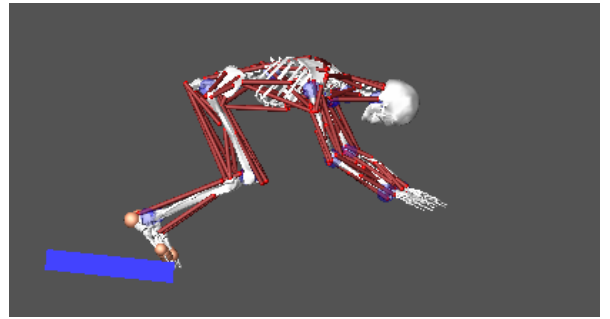

Figure 5. 3D suimmer's body model in the maximal efforts moment of the impulse phase. Time = 0.6013 s.

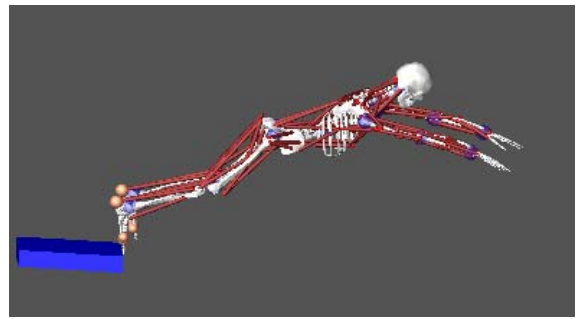

Figure 6. 3D suimmer's body model in the final grab start moment. Time = 0.784 s.

The morphological parameters of the subjects were defined using the measured values for the height and the mass and the anthropometric tables from Dempster et al., 1959. The sum of segments energies was obtained using the equations for the sum of segments energies as defined in Winter (1990).

During the impulse phase, subjects were represented using an open tree structure composed of eight straight segments connected via frictionless joints (Fig. 1). Input data for the model consisted of the fitting angles calculated at each joint (Fig.2 and Fig.3) and the subjects' morphological parameters.

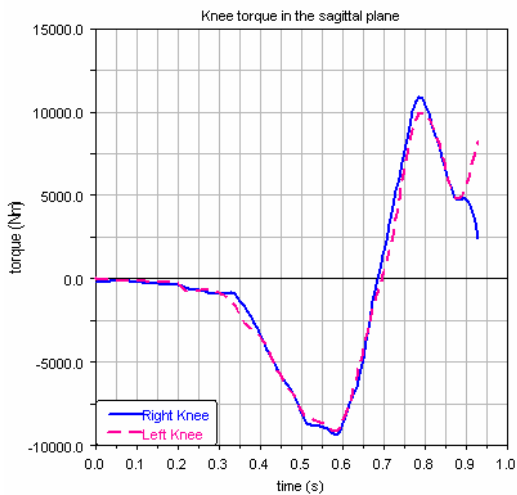

Figure 7. The knee joint torques/time functions (right and left). Time period = 0.00 s ÷ 0.92 s.

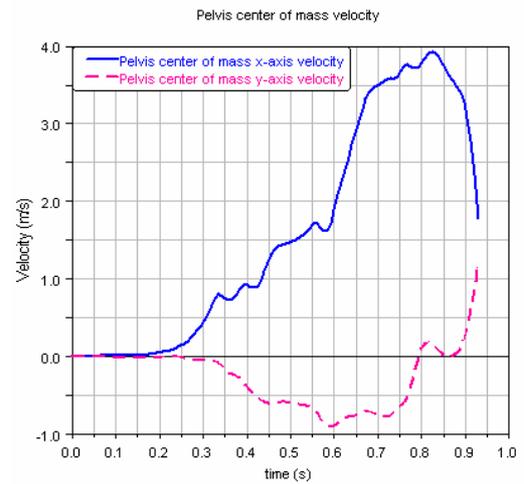

Figure 8. The velocity/time functions of the pelvis centre of mass in the sagittal plane for the time period 0.00 s ÷ 0.92 s

Based on analysis of swimmers' forces and joint moments exerted during the impulse, the model predicts the total body power during the impulse and the position, the initial angle and the velocity of the segments centres of mass at the instant of takeoff.

## III. RESULTS

For every swimmer, 3D model was generated using BRG.LifeMOD$^{TM}$ Biomechanics Modeler on the base of individual morphologic parameters and kinematic data obtained by high speed video camera and passive markers attached at the level of each important articulation. 3D swimmer's body models for four characteristic moments during the grab start are shown: for the initial moment of the impulse phase, t = 0 s (Fig. 4); for the maximal efforts moment, t = 0.6013 s (Fig. 5); for the final start moment, t = 0.8781 s (Fig. 6).

For each joint, the dynamic torque, force and power were determined using the inverse dynamic equations (Winter, 1990; Houel, 2004). The knee joint torques/time functions (right and left) for the time period 0.00 s ÷ 0.92 s are shown on Fig. 7. The velocities of the segments centres of mass were also calculated. The velocity/time functions of the pelvis centre of mass in the sagittal plane for the time period 0.00 s ÷ 0.92 s are shown on Fig. 8 (along the x-axis and the y-axis). The model was validated by comparing the predicted ground reactions forces during the start time period and the measured by AMTI force-plate ground reactions (Fig. 9).

## IV. CONCLUSION

The proposed approach allows predicting swimmer's joint moments for each important articulation during the impulse phase of the grab start and to analyze the segmental coordination for each studied swimmer in order to optimize the performance. The model was successfully validated by comparing the predicted ground reactions during take-off with the experimental ground reactions data collected in situ.

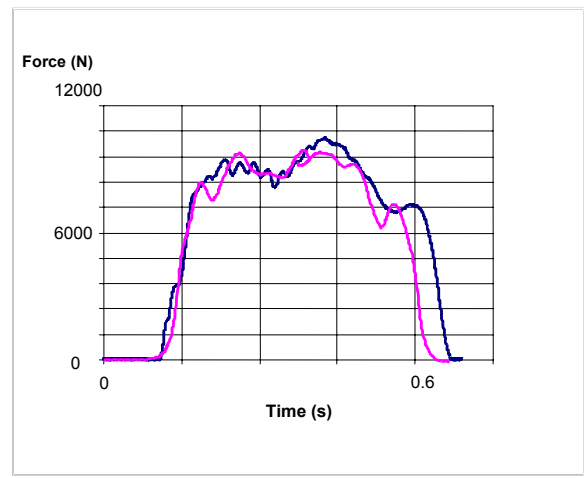

Figure 9. Comparison between the mesured ground reaction forces (bold) and the predicted by the model during the start period of maximal efforts